# A Mass-Conserving 4D XCAT Phantom for Dose Calculation and Accumulation

Christopher L. Williams[1,2], Pankaj Mishra[1,2], Joao Seco[2,3], Sara St. James[1,2], Raymond H. Mak[1,2], Ross I. Berbeco[1,2], John H. Lewis[1,2]

[1]Brigham and Women's Hospital and Dana-Farber Cancer Institute, Boston, MA
[2]Harvard Medical School, Boston, MA
[3]Massachusetts General Hospital, Boston, MA

Email: cwilliams@lroc.harvard.edu , jhlewis@lroc.harvard.edu

**Abstract**
**Purpose:** The XCAT phantom is a realistic 4D digital torso phantom that is widely used in imaging and therapy research. However, lung mass is not conserved between respiratory phases of the phantom, making detailed dosimetric simulations and dose accumulation unphysical. A framework is developed to correct this issue by enforcing local mass conservation in the XCAT lung. Dose calculations are performed to assess the implications of neglecting mass conservation, and to demonstrate an application of the phantom to calculate the accumulated delivered dose in an irregularly breathing patient.
**Methods:** A displacement vector field (DVF) between each respiratory state and a reference image is generated from the XCAT motion model and its divergence is calculated and used to correct the lung density. A series of phantoms with regular and irregular breathing (based on patient data) are generated and modified to conserve mass. Monte Carlo methods are used to simulate conventional and SBRT treatment delivery. The calculated dose is deformed and accumulated using the DVF. Results from the mass-conserving and original XCAT are compared. A 4DCT is simulated for the irregularly breathing patient, and a 4DCT-based dose estimate is compared with the accumulated delivered dose.
**Results:** The presented framework successfully conserves mass in the XCAT lung. The spatial distribution of the lung dose was qualitatively changed by the use of mass conservation; however the corresponding DVH did not change significantly. The comparison of the delivered dose with the 4DCT-based prediction shows similar lung metric results, however dose differences of 10% can be seen in some spatial regions.
**Conclusions:** The XCAT phantom has been successfully modified so that it conserves lung mass during respiration, enabling it to be used as a tool to perform dose accumulation studies in the lung without relying on deformable image registration. Neglecting mass conservation can result in erroneous spatial distributions of the dose in the lung. Using this tool to simulate patient treatments reveals differences between the planned dose and the calculated delivered dose for the full treatment. The software is freely available from the authors.

## I. INTRODUCTION

Respiratory-induced motion is a major source of uncertainty in radiotherapy treatments of thoracic and upper-abdominal lesions. Even in the presence of motion-limiting devices, and image-guided patient setup, intra-fraction anatomic motions of as much as 3 cm or more are observed in clinical practice[1]. This motion deforms and shifts the anatomy, complicating radiotherapy treatment planning, delivery and assessment. The accurate calculation of dose distributions in the presence of this motion is an active area of research[2–7], and remains a significant unresolved issue in radiotherapy.

Techniques for calculating delivered dose in the presence of motion typically use 4-dimensional computed tomography[8] (4DCT) to acquire a series of volumetric images of the patient at different respiratory phases. In current clinical practice, treatment planning and dose calculation often rely on the use of an averaged intensity projection (AIP) with an expanded target volume to account for motion.

However, due to motion, the density of each voxel of the AIP represents a linear combination of different anatomical volumes, possibly from different structures. For example, voxels in the AIP near the lung-diaphragm boundary represent an average of voxels belonging to both the lung and the diaphragm (and organs inferior to the diaphragm). Consequently, dose distributions calculated using an AIP do not accurately represent the true dose delivered to a specific anatomical volume.

Dose distributions can also be computed separately for each 4DCT phase, with deformable image registration (DIR) being used to map the voxels in each image back to a common reference phase[9, 10]. The dose distribution is accumulated using this same mapping. However, this practice of dose accumulation remains controversial, primarily due to its reliance on DIR[11, 12]. The accuracy of DIR algorithms can vary, often depending on the algorithm that is used[13], which can lead to significant uncertainties in the resultant dose distributions[14]. Consequently, the results of DIR-based dose accumulation studies can have large uncertainty, and need additional verification before their results can be fully trusted in clinical practice.

It is difficult to verify the results of DIR-based dose accumulation with phantom measurements because measurements are needed throughout the entire deforming volume. Several groups have recently had success in using deformable gel dosimeters[15, 16, 12] for this purpose (particularly for small deformations). These gels are however limited in the types of deformations they can perform; in particular they have constant electron density and do not yet accurately represent the types of deformations that occur in the lung. Thus, DIR-based dose accumulation still remains largely untested for calculating accumulated lung dose in the presence of respiratory motion.

The lack of verified dose accumulation makes it difficult to assess novel treatment techniques. Various strategies for lung motion management that seek to reduce the dose delivered to surrounding normal tissue have been proposed, however their dosimetric effect has not been rigorously and quantitatively established. A realistic, deformable digital phantom can serve as a useful tool to help answer these questions.

The 4D extended cardiac-torso (XCAT) phantom[17] is a realistic dynamic digital phantom that has been widely used in radiologic and nuclear medicine imaging studies[18–22]. Several studies have also adapted the XCAT (or its predecessor, the NCAT phantom) for use in radiotherapy simulations by combining it with dose calculation algorithms[23, 24]. The XCAT has been adapted to reproduce observed irregular breathing motion[25] and can natively produce the motion vectors between respiratory phases.

A limitation of the XCAT phantom is that it does not conserve the mass of the lung during respiration, making it unrealistic and essentially unphysical to accumulate dose between respiratory phases. McGurk *et al.*[24] addressed this issue in the NCAT (a predecessor of XCAT) phantom by correcting the mean density of the lung in different respiratory phases to match observed patient data. This method conserves mass globally in the lung, but as the authors acknowledged, regional variations in lung density were not considered. Accurately conserving mass (and energy) for these local variations is necessary in order to accumulate dose in different areas of the lung, and consequently a more accurate mass conservation framework is needed.

In this work, we present a method for modifying the XCAT so that it overcomes these limitations, and conserves mass locally, allowing it to be used for dose accumulation studies. We then illustrate the use of dose accumulation and assess the implications of the mass-conserving framework for several test cases.

## II. MATERIALS AND METHODS
### II.A. The XCAT Phantom
The XCAT phantom uses data from the National Library of Medicine Visual Human Project[26] to create a digital model of human anatomy. Each anatomical structure in the phantom is analytically represented by a set of non-uniform rational B-spline (NURBS) surfaces, which are fit to the reference anatomy. The analytical nature of the NURBS model enables the phantom to be deformed or adapted to account for anatomical variation or deformation. The XCAT phantom also includes deformations from respiratory and cardiac motion. The respiratory and cardiac motions are derived from observed 4D tagged magnetic resonance imaging and 4DCT of healthy volunteers, combined with a simple model of respiratory mechanics[17]. The expansion of the pleural space is governed by two parameters: the height of the diaphragm and the anterior-posterior radius of the chest wall. The structures representing the ribcage are rotated upwards and outwards to achieve the desired chest wall expansion, and the diaphragm is contracted downwards during inspiration. The time evolution of these parameters is user-configurable, and can be used to control the respiratory profile. The XCAT phantom software can additionally output a displacement vector field (DVF) defining the motion of each voxel in the phantom as respiratory and cardiac motion occurs. The XCAT DVFs (which we will refer to as XDVFs for clarity) are similar to DVFs calculated with DIR techniques, with the important distinction that the XDVFs are determined directly from the motion of the XCAT NURBS surfaces, and do not rely on any type of DIR. Consequently, the XDVF can be used to unambiguously determine the location and deformation of every voxel in the phantom throughout respiration without the uncertainty introduced by using DIR-derived DVFs.

## II.B. Mass Conservation

In its default configuration, the XCAT software produces 3D grids of x-ray linear attenuation coefficients at specified time steps. Although the anatomic structure of the phantom is deformed with respiratory and cardiac motion, the linear attenuation coefficients, and consequently the density, remain at a fixed value within each anatomical structure. This behavior is problematic for structures that have volumetric changes during respiration, because the total mass of the structure is not conserved. This effect is particularly important in the lungs, where the volume and density are observed in patient images to change by more than 20% over the course of the respiratory cycle[27, 28, 1].

In this work, we have developed a method to enforce a local conservation of mass in the XCAT phantom by using the XDVF relative to its initial reference phase (*e.g.,* full exhale) to correct the phantom at each other phase (referred to as "target" phases). The density of each individual voxel is changed based on the deformation of that specific voxel. This process is complicated by the fact that there is not a one-to-one correspondence between voxels in the reference phase and target phase (due to the changing lung volume and finite voxel dimensions). For the studies performed in this work, we use a full exhale phase as the reference phase.

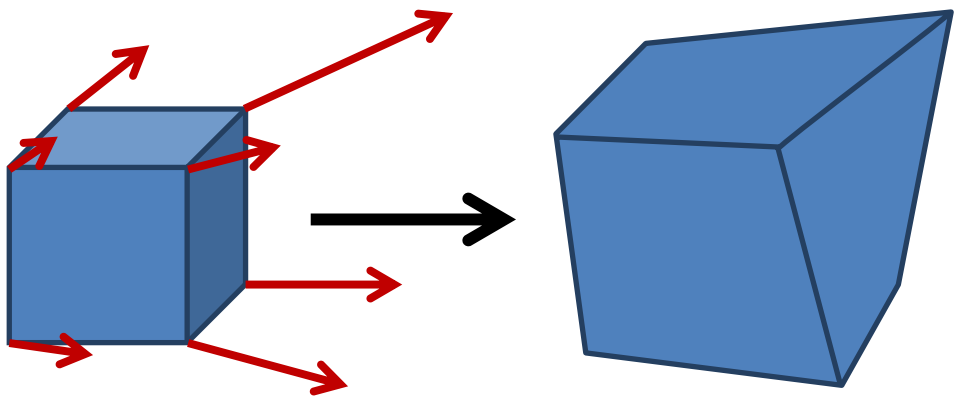

FIG. 1. Cartoon illustration of how the XDVF motion vectors (red arrows) are used to deform a rectangular virtual voxel into a hexahedron. The volume change between the initial virtual voxel and the deformed hexahedron is used to quantify the local expansion or contraction of the phantom due to the motion described by the XDVF.

We first determine how the volume of each voxel in the reference phase changes when deformed to the target phase using the XDVF. In the reference phase, we define a set of "virtual voxels," using the centers of each collection of 8 adjacent voxels as the vertices. The XCAT motion vectors are then used to deform the corners of these virtual voxels and to then calculate the change in the virtual voxel volume relative to the initial reference phase voxel volume (a cartoon of this process is shown in FIG. 1). Under deformation, each initially rectangular voxel maps to a hexahedron in the target phase which we then subdivide into 6 component tetrahedrons in order to calculate the volume. The resulting change in volume between the original rectangular virtual voxel and the deformed hexahedron represents the local volume change of each reference phase voxel under the specified XDVF. This is essentially a discrete calculation of the local divergence or the determinant of the Jacobian of the XDVF. These volume changes are calculated for each virtual voxel in the reference phase of the phantom. Under the assumption that the mass in each voxel remains unchanged during the deformation, the density variation induced by the motion is inversely proportional to the relative volume change. An example coronal slice of the calculated voxel volume change during respiration is shown in FIG. 2.

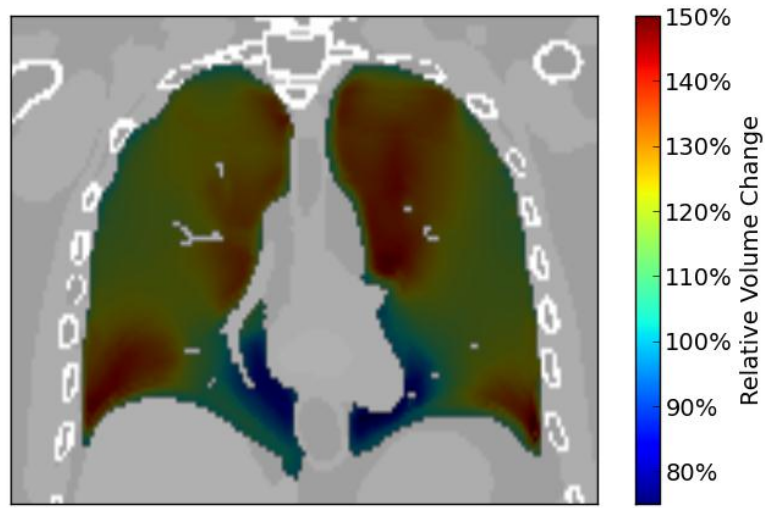

FIG. 2. Example coronal slice showing the relative volume change in different areas of the lung between an inhaled phase and a reference phase at full exhale using the XCAT motion vectors. The lung is shown at full inhale, with the color scale showing how the volume of in each area of the lung has changed relative to the volume at full exhale.

In order to correct the density of the phantom in the target phase, we next determine the location in the reference phase phantom associated with each target phase voxel. We employ the fixed-point method of Chen *et al*[29] with linear interpolation to determine the inverse vector field that maps each voxel in the target phase back to the reference phase (i.e. the inverse of the XDVF). The density of each voxel in the target phase is then determined from the reference phase density deformation map at the position calculated from the inverted XDVF.

Due to the voxelized nature of the XDVF, the calculated local volume changes occasionally exhibit defects at the boundaries between anatomical structures. To mitigate these effects, we apply a 3x3x3 voxel median filter to the determined voxel volume deformations. These voxelization effects are more pronounced near small structures, where a large fraction of the voxels is near the boundary. Consequently, for this study we only correct the lung density of the phantom.

## II.C. Tumor-Lung Synchronization

To simulate a realistic lung cancer patient, a lesion was inserted into the XCAT phantom so that it moved synchronously with the chest wall, diaphragm and surrounding lung tissue. The issues of chest wall and diaphragm synchronization in the XCAT were addressed by Mishra *et al*.[25] The position of the tumor is controlled indirectly by setting the motion of the diaphragm and chest wall (cardiac motion can also influence the tumor position, but its effect is minimal for the right-sided tumors simulated in this study).

The combination of a rigid tumor and deforming surrounding lung tissue can create an unphysical situation where regions of lung appear or disappear in the area immediately surrounding the tumor during respiration, invalidating dose tracking in this region. To correct this issue we maintain the boundary between the tumor surface and the lung tissue. This introduces a small deformation to the tumor volume, as it deforms in the same way as the surrounding lung. In reality the more rigid tumor body would probably deform less than the surrounding lung[30]. This is a limitation of the current implementation of lung lesions in the phantom software.

## II.D. Phantom Generation

We employed the methods described in Sections II.B and II.C to generate three sets of digital phantoms. The phantoms were produced with 2 mm voxel widths in the LR and AP direction, and a 2.5 mm voxel width in the SI direction. Phantoms were generated with small (2 cm) and large (6 cm) spherical lesions in the right lung. Both of these phantoms used regular breathing with the standard XCAT breathing profile (almost sinusoidal), and a respiration period of 5 s. The maximum tumor motion was 1.2 cm in the cranial-caudal direction. Individual phantom phases were output every 200 ms. These phantoms were generated both with and without mass conservation enabled, and were used to assess the impact of mass conservation on the calculated dose.

| Phantom Set | Lesion Diameter | Number of Time Steps | Time per Step | Tumor Motion Type |
|---|---|---|---|---|
| Regular SBRT | 2 cm | 24 | 200 ms | Regular |
| Regular Conventional | 6 cm | 24 | 200 ms | Regular |
| Irregular Breathing | 2 cm | 67 | 1 s | Recorded Patient |

TABLE I. Summary of phantoms generated and used in this study. Versions of the regularly breathing SBRT phantom and the regularly breathing conventional phantom were produced both with and without mass conservation, while the irregularly breathing phantom was only generated in a mass conserving version.

An additional phantom with a 2 cm spherical tumor was generated to simulate a patient with an irregular breathing pattern. The tumor track for this phantom was derived from data obtained by Berbeco *et al*.[31] using a fluoroscopic marker tracking system[32] at the Nippon Telegraph and Telephone Corporation Hospital in Sapporo, Japan. A 2 cm tumor with this motion was inserted in the lung and synchronized with the diaphragm, chest wall and surrounding lung tissue using the procedure described in Section II.C. A total of 67 s of breathing data from a single patient trace was used, with time steps produced every 1 s (due to computational constraints). This phantom was generated with mass conservation enabled, and was used to demonstrate an application of the mass conserving phantom. A summary of the phantoms that were generated is displayed in TABLE I. Sample coronal slices from these phantoms are shown in FIG. 3, and the breathing trace from the irregularly-breathing patient is shown in FIG. 4.

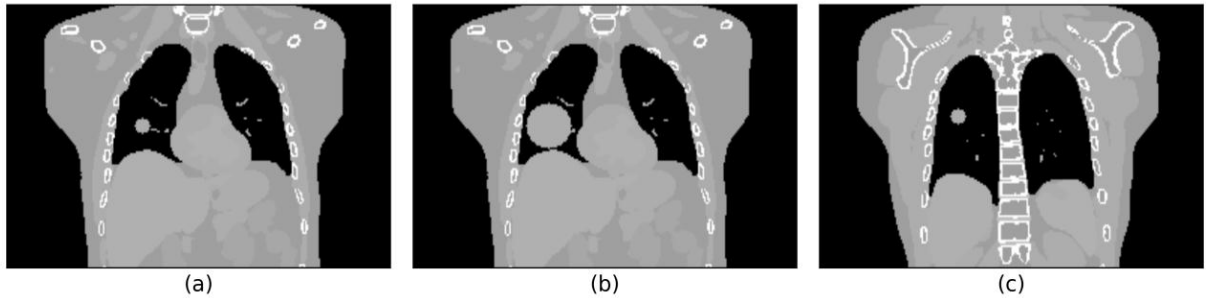

FIG. 3. Sample coronal slices from the three digital phantoms generated for this study. Panel (a) shows the regularly-breathing phantom with a 2 cm tumor, panel (b) shows the regularly-breathing phantom with a 6 cm tumor, and panel (c) shows the phantom generated based on the irregular motion patient breathing trace. The slices are chosen to pass through the center of the lesions (right lung).

## II.E. Treatment Planning

Simulated treatments were planned for each of the digital phantoms to evaluate the dose received by both the tumor and the lungs. Planning imaging was simulated by calculating the average intensity projection (AIP) and maximum intensity projection (MIP) for each of the phantoms. The AIP and MIP were based on one full breath for the regular breathing phantoms, and were based on the first 6 s of imaging for the irregularly breathing phantom (~2 breaths). This method approximates the 4DCT simulation protocol used in our clinic. The AIP was used for dose calculations during treatment planning and the MIP was used to define the internal target volume (ITV).

A stereotactic body radiation therapy (SBRT) treatment was planned for each of the phantoms with 2 cm lesions. Treatment plans based on an arrangement of 9 non-coplanar static MLC fields frequently used in our clinic were developed. A PTV was defined using a 5 mm expansion of the ITV. The collimator jaws and MLC leaves were fit to the PTV in the beams-eye-view for each field. The plan was normalized to the 82% isodose level so that a dose of 54 Gy covered 95% of the PTV, with a typical clinically acceptable dose distribution. A conventional treatment was planned for the regularly breathing phantom with the 6 cm lesion. A 4-field treatment plan with 4 sub-fields was used (AP-PA with obliques). This plan was normalized to the 86% isodose level so that a dose of 66 Gy covered 95% of the PTV.

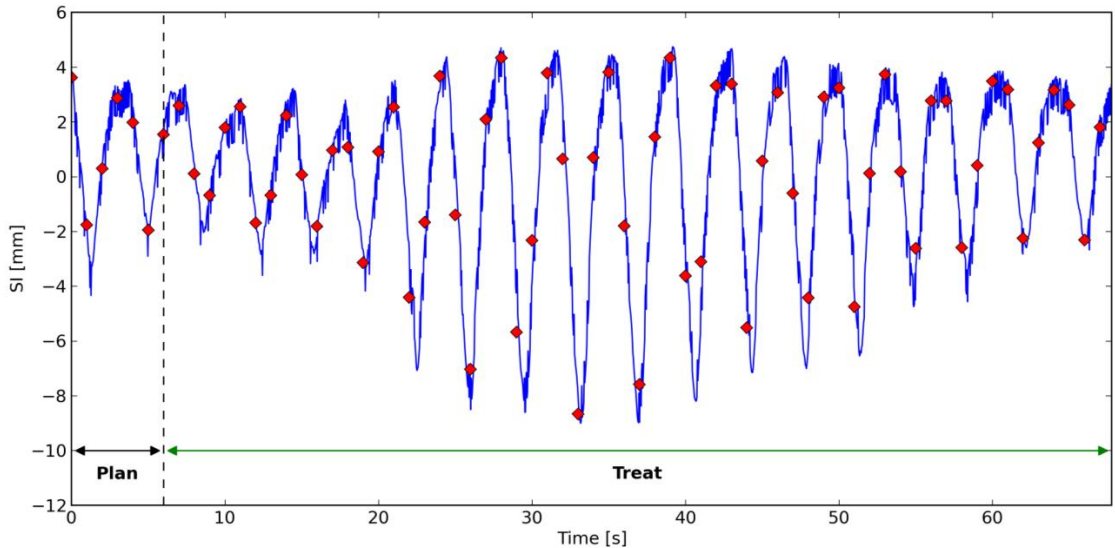

FIG. 4. Superior-inferior (SI) component of the patient breathing trace used to generate the irregular-breathing phantom used in this study. The solid (blue) curve is the measured patient breathing motion. The diamonds (red) show the SI position of the tumor centroid calculated for each time step of the phantom. The time intervals used for treatment planning and treatment simulation using this phantom are also indicated.

## II.F. Dose Calculation and Accumulation

In order to calculate the dose delivered to the phantom in a simulated treatment, we first calculate the instantaneous dose at each time step in the generated phantom, and then map this dose back to a reference phase of the phantom (we use a full exhale phase in this study). The simulated delivered dose for each treatment plan was calculated using a well-validated Monte Carlo dose engine developed by Seco *et al.*[4] based on the DPM[33] and EGSnrc[34] software packages. The dose was calculated separately for each beam and each time step generated for the phantoms. The particle phase space was computed so that it contained at least $10^7$ particles in each field beyond the MLC, ensuring a statistical uncertainty of less than 2% in the high-dose region for each beam and time step.

We calculate the dose to each phantom time step by simulating the delivery of each beam/subfield. These individual dose contributions are scaled by the number of monitor units (MUs) planned for that beam/subfield and summed. The same number of MUs is used for both the mass conserving and original XCAT phantom simulations. The XDVF is then used to map the dose distribution back to the reference phase by interpolating the dose delivered to the displaced locations of the reference phase voxel centers. The reference phase doses from each phantom time step are then averaged to produce the full accumulated dose distribution. The delivered dose was assessed by delivering the plans developed in Section II.E to the mass conserving and original XCAT phantoms.

Two other calculations were performed for comparison with the accumulated delivered dose: 1) dose was calculated on the AIP to represent current common clinical practice; and 2) dose was calculated on a 10-phase 4DCT (generated from the same data used to define the AIP and MIP) and XDVFs were used to accumulate the dose to the full exhale phase. This second "4DCT-based dose" method is meant to represent a state-of-the-art calculation that incorporates all of the motion information in 4DCT[10, 35, 36].

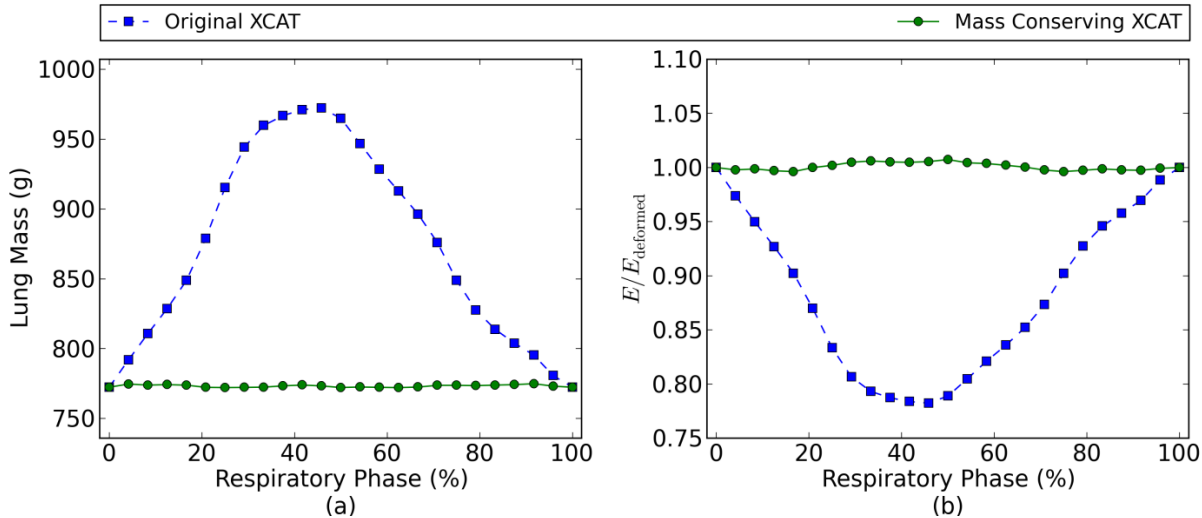

FIG. 5. Panel (a) compares the lung mass as a function of respiratory phase in the original XCAT phantom with the lung mass after the implementation of the mass conservation scheme described in Section II.C. Panel (b) compares the ratio of energy in the dose distribution before and after deformation to a reference respiratory phase (full exhale) for the conventional radiation plan. The change in energy seen in the original XCAT phantom results from the lack of mass conservation as the chest expands during breathing. The residual energy- and mass- conservation defects in the mass conserving XCAT results are due to voxel size effects.

## III. Results

### III.A. Mass and Energy Conservation

The conservation of energy under dose deformation is an important internal consistency check of our phantom mass conservation and dose accumulation procedure. For the type of deformations that occur during respiration, deforming a dose distribution between phases of the phantom should not change the total amount of energy contained in that dose distribution. We assessed energy conservation by evaluating:

$$E = \sum_n D_n \rho_n \Delta v, \qquad (1)$$

where $E$ is the total energy deposited in the phantom, $D_n$ is the dose deposited in voxel $n$, $\rho_n$ is the mass density in voxel $n$, $\Delta v$ is the voxel volume and the sum is taken over all voxels in the lung of the phantom. We computed the energy separately for the dose delivered at each respiratory phase in the regularly-breathing phantoms. We then deformed the dose back to the full exhale phase phantom and recalculate the energy. For the phantom and deformation to be self-consistent, the total energy in the dose distribution should not be changed by deforming between respiratory phases. An example plot of the ratio between the energy before and after the deformation for the conventional radiation plan is displayed in FIG. 5, showing that energy is properly conserved under deformation when using the mass-conserving XCAT phantom. There maximum residual defects in the mass and energy conservation are 0.3% and 0.7% respectively. These small conservation errors arise from the voxelized nature of the density correction procedure, and are reduced if a higher resolution phantom is used.

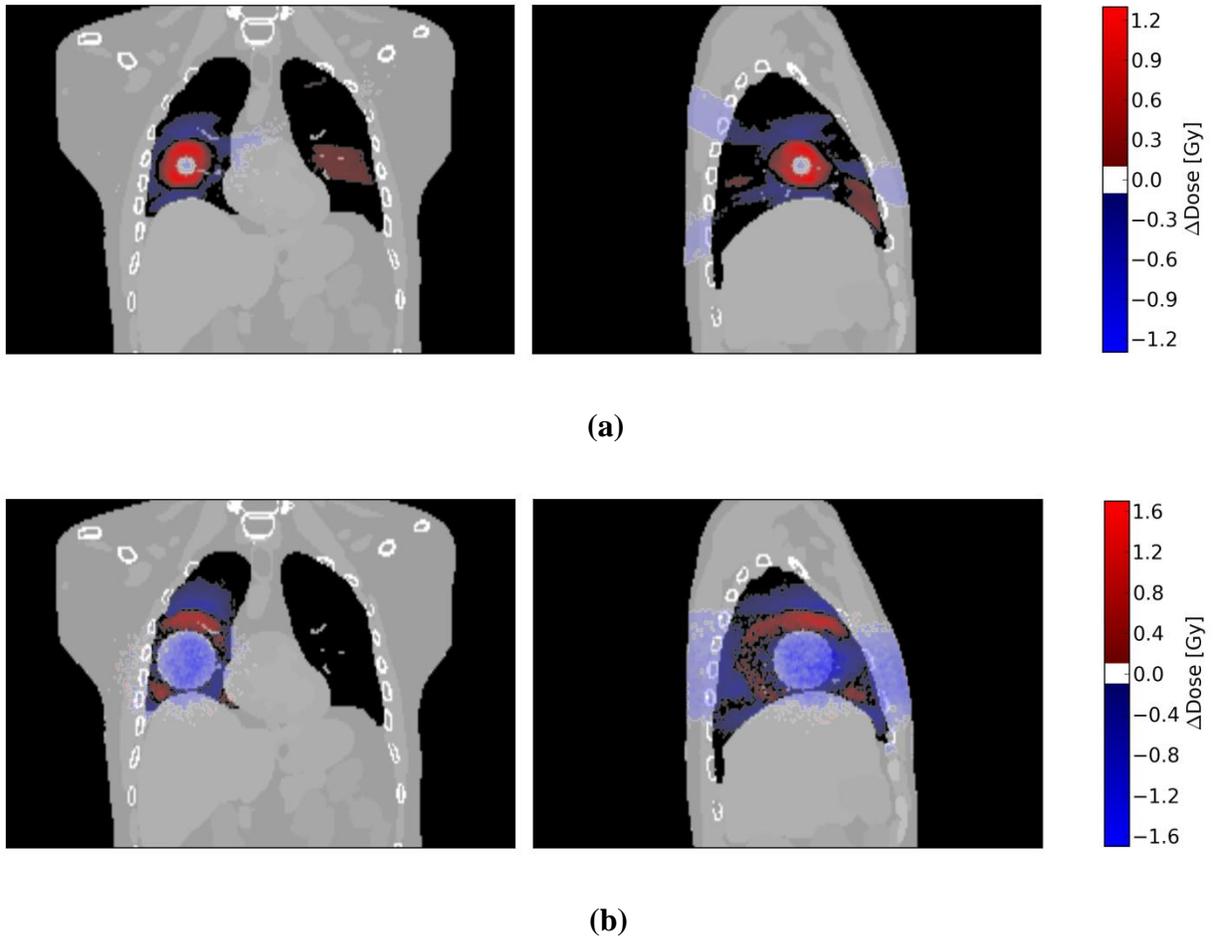

FIG. 6. Dose distribution differences between the mass conserving and original XCAT phantoms for the 54 Gy SBRT treatment (a) and the 66 Gy conventional treatment (b) with regular breathing. The dose difference, **Δ**Dose, is defined as the dose calculated in the original XCAT phantom minus the dose calculated using the mass conserving phantom. The coronal and sagittal slices are chosen to go through the center of the target lesion. The red shading corresponds to regions where a dose calculation using the original XCAT would overestimate the dose, and blue shaded areas are regions where the original XCAT would underestimate the delivered dose relative to the mass-conserving phantom. Regions with dose differences smaller than 0.1 Gy in magnitude are not shaded.

### III.B.    Dose Distribution Changes

A comparison of the dose distributions produced from the simulated treatments show differences in the overall dose distribution depending on whether the mass conserving phantom is used. The largest voxel dose difference is 1.4 Gy for the SBRT treatment and 1.5 Gy for the conventional treatment. Coronal and sagittal slices through the dose distributions are shown in FIG. 6. In both cases, the dose calculated using the original XCAT slightly underestimates the dose delivered to the tumor relative to the mass conserving XCAT. There is also a region immediately surrounding the tumor in both cases where the original XCAT overestimates the dose to the surrounding lung tissue.

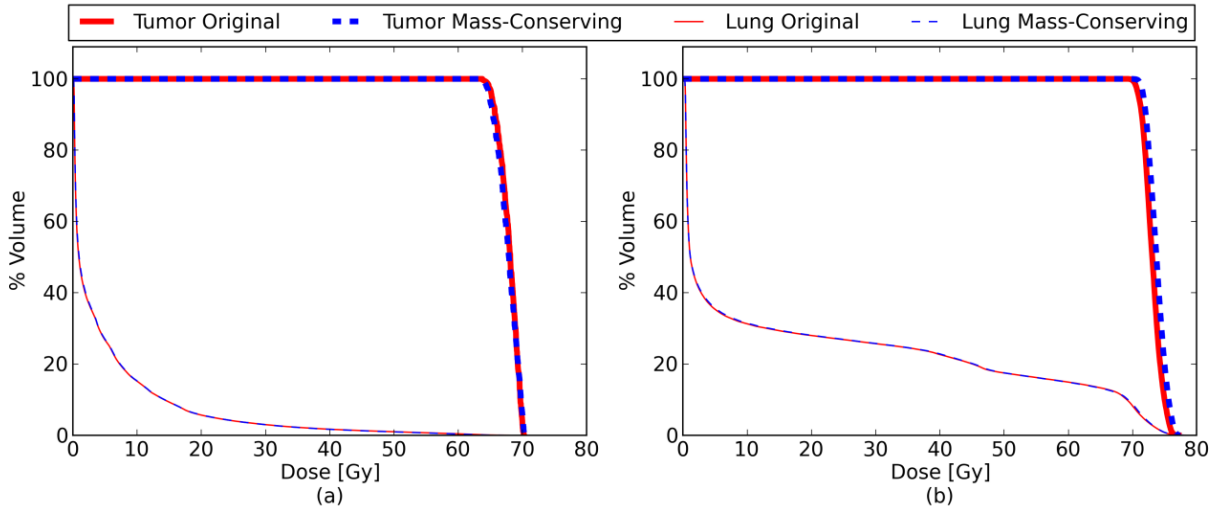

FIG. 7. Dose volume histograms for the tumor (thick) and lung (thin) for the same treatment plan delivered over one complete breathing cycle to the original (blue, dashed) and mass-conserving (red, solid) XCAT phantoms. Panel (a) shows the result for the SBRT treatment, while panel (b) shows the conventional treatment. The histograms for the original and mass-conserving lung dose are nearly indistinguishable.

### III.C. Dose Metric Changes

A comparison of the dose volume histograms (DVHs) for the lung and tumor are shown in FIG. 7. Common metrics for these dose distributions are presented in TABLE II. Because this is a digital phantom, no contouring was necessary to compute the DVHs of particular anatomic structures. The DVHs from the original XCAT and the mass conserving XCAT do not appear to have any clinically significant difference, although minor variation can be seen in the dose delivered to the tumor.

| Plan | Phantom | Tumor | | Lung | | |
|---|---|---|---|---|---|---|
| | | D95 [Gy] | Max [Gy] | V20 | V5 | Mean [Gy] |
| SBRT | Original | 69.5 | 70.2 | 5.7% | 27% | 4.9 |
| SBRT | Mass Cons. | 69.5 | 70.3 | 5.7% | 27 % | 4.9 |
| Conventional | Original | 76.1 | 77.2 | 28% | 36% | 17.3 |
| Conventional | Mass Cons. | 76.7 | 77.4 | 28% | 35% | 17.2 |

TABLE II. Dose metrics for the accumulated delivered dose calculation using regularly breathing XCAT phantoms. The conventional and SBRT plans were evaluated using both the original and mass-conserving versions of the XCAT phantom.

### III.D. Respiratory Phase Dependence of Delivered Dose

In order to assess which respiratory phases show the largest effect from the addition of mass conservation, we simulated delivering a treatment to each respiratory phase, and then deformed the dose back to a common reference phase. This experiment could also be important for respiratory gating or breath-hold treatments. The results of this experiment are shown in FIG. 8. The lung DVH curves are similar for the original and mass-conserving phantoms, showing decreases slightly during the inhale respiratory phase, particularly at high dose levels. The lung V70 changes from 10% at exhale to 7% at inhale (after mapping the dose back to a common lung exhale volume). When comparing the original XCAT tumor DVH to the mass-conserving tumor DVH, the inhale D95% changes from 67.2 Gy to 67.7 Gy, and the inhale D50% changes by 1 Gy, from 72.2 Gy to 73.2 Gy. In general the mass-conserving XCAT shows smaller changes in the DVH curves as a function of respiratory phase than the original XCAT phantom.

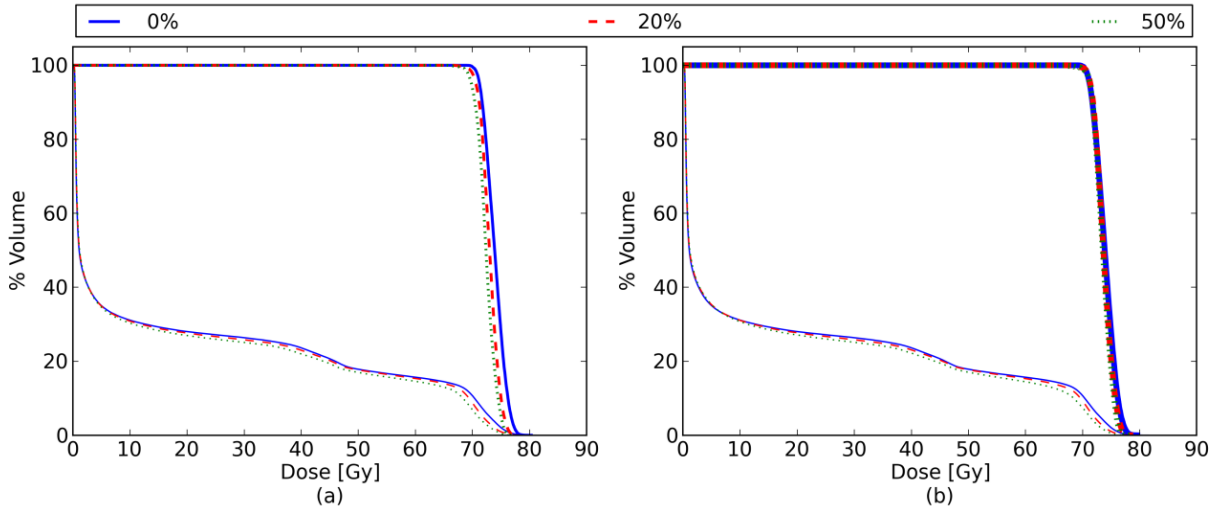

FIG. 8. DVHs as a function of respiratory phase for the regularly breathing phantom. The DVHs for the tumor are shown as thick lines and the lungs as thin lines. Panel (a) shows the result for the original XCAT, while panel (b) shows the results for the mass-conserving XCAT. The breathing phases shown vary from 0% (full exhale) to 50% (full inhale).

### III.E.  Realistic Patient Breathing

The accumulated dose delivered to the irregularly breathing phantom is shown in FIG. 9, along with the DVHs calculated on the AIP used for treatment planning. The delivered D95 to the tumor is 60.9 Gy, in comparison to the planned D95 to the PTV of 54.0 Gy. The PTV dose calculated on the AIP is smaller than the actual dose delivered to the tumor for this simulated patient, due to the low normalization used in the SBRT plan, and the tumor motion.

Lung DVHs are shown for both the accumulated delivered dose and the AIP prediction. For this comparison, the ITV has been subtracted from the lung. Because the volume and location of the lung changes during respiration, the delivered lung dose was evaluated both at maximum exhale and maximum inhale to provide a comparison to the AIP evaluation. A comparison of dose metrics is presented in TABLE III.  The lung DVH curves are in general similar between the AIP and both delivered dose curves, although the AIP calculation appears to slightly underestimate the volume of lung in the low dose region ($< 5$ Gy) for this patient.

| Dose Calculation | Tumor/PTV | | Lung-ITV | | |
|---|---|---|---|---|---|
| | D95 [Gy] | Max [Gy] | V20 | V5 | Mean [Gy] |
| AIP | 54.0 (PTV) | 65.9 (PTV) | 2.7% | 23% | 4.0 |
| Accumulated Delivered Inhale | 60.9 (Tumor) | 65.4 (Tumor) | 3.9% | 25% | 4.4 |
| Accumulated Delivered Exhale | | | 3.2% | 25% | 4.4 |

TABLE III.  Dose metrics comparing the AIP-based plan and the accumulated dose for the realistically breathing phantom.  The target dose is evaluated for the PTV in the AIP-based dose distribution and for the actual tumor volume in the accumulated delivered dose distribution.  The Lung-ITV dose is evaluated at both inhale and exhale for the accumulated delivered dose distribution.

DVHs based on the 4DCT dose calculation show no appreciable differences when compared to the accumulated delivered dose DVHs. A comparison of dose metrics of these two distributions is shown in TABLE IV. These metrics and DVHs are in better agreement with the delivered dose than the AIP-based metrics are. The spatial differences in the dose distribution calculated from 4DCT compared to the delivered dose are shown in FIG. 10. These slices show 5 Gy differences in the dose distributions, with

Normal Tissue Dose Calculation with a Mass-Conserving 4D XCAT Phantom

the 4DCT dose calculation predicting greater dose in the lower regions of the lung and reduced dose in the upper regions.

| Dose Calculation | Tumor | | Lung | | |
|---|---|---|---|---|---|
| | D95 [Gy] | Max [Gy] | V20 | V5 | Mean [Gy] |
| 4DCT | 61.0 | 65.4 | 4.2% | 25% | 4.4 |
| Accumulated Delivered | 60.9 | 65.4 | 4.1% | 25% | 4.4 |

TABLE IV. Dose metrics comparing the 4DCT dose calculation and the accumulated dose for the realistically breathing phantom.

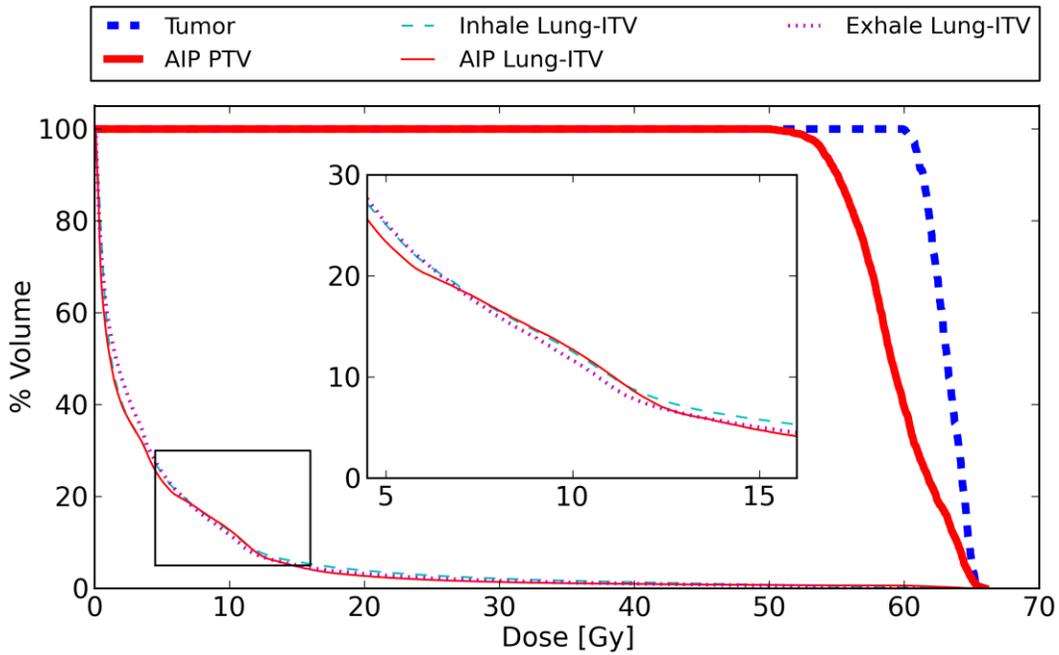

FIG. 9. Delivered dose compared to planned dose for the mass-conserving XCAT phantom generated using an observed patient breathing trace. The planned dose is based on the conventional ITV method calculated on the AIP. The lung DVH is calculated for the region outside the ITV, and is assessed at both the patient's peak exhale and peak inhale (because the lung volume is different for both of these phases). The inset shows a blowup of the lung dose region between 5 and 15 Gy.

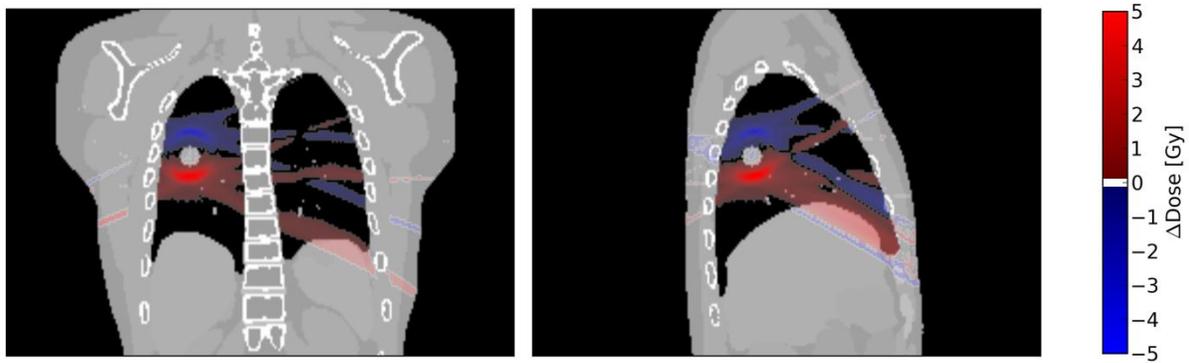

FIG. 10. Sagittal and coronal slices of the dose distribution difference between the 4DCT dose calculation and the accumulated delivered dose for the patient-based XCAT phantom with a 54 Gy SBRT plan. Regions where the 4DCT dose calculation predicts more dose than is delivered are shown in red, and regions where the 4DCT dose calculation predicts less dose than is delivered are shown in blue.

## IV. DISCUSSION

The mass-conserving XCAT provides a realistic digital phantom which allows the physically accurate calculation and accumulation of dose. Combined with the XCAT's ability to reproduce observed irregular breathing motion, it can serve as a useful tool for realistically assessing motion management techniques and dose accumulation in radiotherapy independent of any DIR techniques.

In this work, we have demonstrated this capability by comparing the planned and delivered dose distributions for an irregular breathing patient using a conventional ITV-based plan. This provides one example of how the mass-conserving XCAT can be used to assess the efficacy of different treatment planning and delivery strategies in the presence of motion. In future work, we plan to use this phantom to further assess the efficacy of motion management techniques such as gating and tracking.

One drawback of this technique is that currently lesions in the XCAT are inserted simply by tagging lung voxels and changing their attenuation coefficients. As discussed in Section II.C, this requires the tumor to deform in the same manner as the surrounding lung tissue, leading to potentially unrealistic deformation of the tumor. Incorporating an additional separate NURBS surface for the tumor within the XCAT phantom software could allow for more accurate tumor representation, while still maintaining the ability to sensibly deform dose in the region around the tumor.

We also note that because the method for mass conservation presented in this work uses the voxelized vector field and attenuations produced from the XCAT software, it is subject to discretization errors, particularly near the boundaries of structures. This may lead to erroneous density values in regions of the phantom where the density is changing rapidly, and is the cause of the small imperfections in the mass and energy conservation seen in FIG. 5. This effect could be ameliorated by incorporating local mass conservation directly into the XCAT software by analytically calculating the local divergence of the deformations and correcting the mass internally before a gridded version of the phantom is produced. Voxelization errors in the dose calculation may also be reduced by using a Monte Carlo technique that intrinsically accounts for the deformation[37, 38].

In the irregular breathing example presented in this study, the dose accumulation process does not account for the serial nature of the beam delivery, the different delivery times of beams, or any time gaps in the delivery due to gantry/couch repositioning. The interplay between these effects could be important and should be investigated further.

The volumetric expansion information derived from the XDVF and used to correct the lung density (such as that displayed in FIG. 2) provides a ventilation image for the XCAT lung[39]. This information could be used in treatment planning investigations that seek to spare the most functional areas of lung during radiotherapy.

One interesting result of this study is the nature of the dosimetric errors introduced by the lack of mass conservation in the original XCAT in the specific case that we investigated. Even though the lung mass increases by ~25% in the original XCAT, the overall effect on the statistics of the accumulated dose distribution is significantly smaller, because the energy deposited increases by nearly the same amount, resulting in a similar overall dose. The most significant changes are in the spatial distribution of the dose, likely as a result of the differing attenuations of the incident x-rays. Deformable registration methods that do not conserve mass could potentially show similar types of errors.

As expected, the DVH metrics calculated from the 4DCT dose were much closer to the delivered dose metrics than those calculated based on the AIP. However, the spatial distribution of the 4DCT dose when compared to the delivered dose was substantially different. This spatial change in the dose distribution could have implications for the use of techniques such as dose painting[40,41], where a specific spatial dose distribution is desired.

## V. CONCLUSION

The addition of mass conservation addresses a key deficiency in the XCAT phantom for radiotherapy applications, making it a suitable platform for dose calculation and accumulation studies. Motion vectors derived directly from the XCAT can be used to deform the dose, thus eliminating any uncertainty introduced by the use of deformable registration techniques and allowing an assessment of the accumulated dose to the lung. The flexibility and analytic nature of the XCAT allows for a wide range of tumor sizes and motions to be evaluated, including matching patient-observed irregular breathing. The accumulated delivered dose distributions generated from Monte Carlo treatment simulations yield realistic results, and highlight the impact that irregular breathing may have on the distribution of dose in the lung. Software to modify XCAT phantoms to conserve mass, and tools to perform dose accumulation are available and can be freely obtained by contacting the authors.

**ACKNOWLEDGEMENTS**

The authors would like to thank Dr. Seiko Nishioka of the Department of Radiology, NTT Hospital, Sapporo, Japan and Dr. Hiroki Shirato of the Department of Radiation Medicine, Hokkaido University School of Medicine, Sapporo, Japan for sharing the patient tumor motion dataset with us. The project described was supported by Award Numbers RSCH1206 (JHL) from the Radiological Society of North America and R21CA156068 (RIB) from the National Cancer Institute. The content is solely the responsibility of the authors and does not necessarily represent the official views of the National Cancer Institute or the National Institutes of Health.

Normal Tissue Dose Calculation with a Mass-Conserving 4D XCAT Phantom

-17-